\documentclass[pra,a4paper,notitlepage]{revtex4-1}
\pdfoutput=1

\usepackage[verbose=true]{microtype}
\usepackage[utf8]{inputenc}
\usepackage[intlimits]{amsmath}
\usepackage{amssymb,amsfonts,amsthm}
\usepackage{enumerate}
\usepackage[capitalise]{cleveref}
\usepackage{graphicx}

\theoremstyle{plain}

\DeclareMathOperator{\sech}{sech}
\renewcommand{\epsilon}{\varepsilon}

\newcommand{\1}{{\openone}}

\newcommand{\be}{\begin{eqnarray}}
\newcommand{\ee}{\end{eqnarray}}
\newcommand{\ba}{\begin{array}}
\newcommand{\ea}{\end{array}}

\newcommand{\ben}{\begin{enumerate}}
\newcommand{\een}{\end{enumerate}}
\newcommand{\bi}{\begin{itemize}}
\newcommand{\ei}{\end{itemize}}

\begin{document}

\title{Construction of the raising operator for Rosen-Morse eigenstates in terms of the Weyl fractional integral}

\author{F.~L.~Freitas}
\address{Instituto Tecnol\'{o}gico de Aeron\'{a}utica, 12228-900 S\~{a}o Jos\'{e} dos Campos, SP, Brazil}

\begin{abstract}
The raising operator relating adjacent bound states for the general, non-symmetric Rosen-Morse potential is constructed explicitly. It is demonstrated that, in constrast to the symmetric (modified P\"oschl-Teller) potential, the operator is non-local and must be expressed applying techniques from fractional calculus. A recurrence relation between adjacent states is derived applying the Weyl fractional integral, which, in contrast to standard recurrence relations, allows the efficient numerical computation of the coefficients of all Jacobi polynomials necessary for the evaluation of the bound state wave functions, providing an application of fractional calculus to exactly solvable quantum systems. \end{abstract} 

\maketitle

\section{Introduction}
\label{sec:introduction}

The Rosen-Morse potential, given by

\begin{equation}
V(x) = -\frac{\hbar^2\alpha(\alpha+1)}{2m\Delta^2}\sech^2\left(\frac{x}{\Delta}\right) + \frac{\hbar^2\beta}{m\Delta^2}\tanh\left(\frac{x}{\Delta}\right),
\label{eq:rosen-morse}
\end{equation}
was considered in early studies of quantum mechanics\cite{Rosen1932} as a model for polyatomic molecules. The first term in the potential has the shape of a finite well, whose width and depth are controlled by the parameters $\Delta$ and $\alpha$, respectively. In addition to that, the second term describes a smooth barrier, with thickness and height determined by the parameters $\Delta$ and $\beta$.

Nowadays, the potential described by \eqref{eq:rosen-morse} has been used in numerous applications. Specifically, its symmetric form, with $\beta=0$ is also referred to in the literature as the modified P\"oschl-Teller potential\cite{Flugge1994}, and has the unusual property of being reflectionless for scaterring states of any energy\cite{Lekner2007} when the parameter $\alpha$ is an integer. Additionally, the Rosen-Morse potential is shape invariant with respect to a supersymmetric (SUSY) transformation, and its exact solution can be obtained in terms of SUSY QM factorization\cite{Cooper2001}. It has been applied to the description of fermions in the early universe, providing a possible explanation for baryogenesis\cite{Prokopec2013}.

Perhaps its most striking modern application is the description of Fermionic modes in topologically trapped vacua. As noted by Jackiw and Rebbi\cite{Jackiw1976}, and then employed in the context of condensed matter systems by Haldane\cite{Haldane1988}, the coupling of Fermi fields to soliton background solutions presents chiral zero modes which violate parity symmetry. For a certain choice of the self-interaction potential for the scalar field in the Jackiw-Rebbi lagrangian, it is possible to reduce the computation of the Fermion masses in the kink background to the solution of the potential \eqref{eq:rosen-morse}\cite{Chu2008}, whose analytical solution implies that the entire mass spectrum can be determined exactly.

The simple determination of the bound state spectrum has been exploited to describe many realistic physical systems of enormous relevance. The Rosen-Morse potential forms the basis of the SSH model\cite{Su1980}, where the scalar field responsible for symmetry breaking is the phonon field of polyacetylene, whose ground state can be forced into a topologically trapped soliton by doping, giving rise to the Jackiw-Rebbi zero mode responsible for the outstanding conductivity of this polymer. Shortly after, the model was generalized for extra dimensions\cite{Takayama1980} and studied in the context of surface physics by Volkov and Pankratov to the study of quasiparticles in the presence of gap inversion\cite{Volkov1985,Pankratov1991} and rediscovered as a useful tool to describe the surface states in topological heterojunctions (THJs)\cite{Inhofer2017,Tchoumakov2017,Freitas2017} and the Fermi arc states in Weyl semimetals\cite{Araki2016}.

Nevertheless, despite its importance, mathematical treatments of the Rosen-Morse eigenstates have been rare in the literature. One possible reason is the outstanding complexity of the calculations involved. In fact, even the determination of the normalization constants for the wave functions requires a lot of work\cite{Nieto1978,Kleinert1992}. With the emergence of topological materials as a well established research program, this lack of results presents serious challenges. First, the simplest treatment of the phenomenology of THJs assumes the symmetric form of \eqref{eq:rosen-morse}, which implies that both the trivial and topological insulators have the same gap. This does not happen for most realistic systems\cite{Inhofer2017}, and therefore the theoretical description is rather limited.

Second, it is expected that the surface potential in smooth topological transitions should deviate reasonably from \eqref{eq:rosen-morse}, and without a good knowledge of the bound states, it is not possible even to calculate perturbative corrections to their energy. This work is a first step in the efficient computation of matrix elements between the eigenstates of the Rosen-Morse potential. The chosen approach is the explicit construction of ladder operators for the bound states\cite{Dong2002}, which was proven to yield efficient recurrence relations in the symmetric P\"oschl-Teller case\cite{Zuniga1996}. The ladder operators necessary for this procedure are fundamentally different from those obtained from SUSY QM, which link states between different Rosen-Morse potentials for different values of $\alpha$, while we are interested in ladder operators relating states of the same potential.

It is the goal of the present article to show that the construction of the raising operator for the non-symmetric ($\beta\neq0$) Rosen-Morse potential is fundamentally different from the symmetric case, already worked out in the literature\cite{Dong2002,Zuniga1996}. The reason for this is that a relation involving different Rosen-Morse eigenstates involves a non-trivial change of parameters in Jacobi polynomials. Nevertheless, it is possible to apply an idea of Ismail\cite{Ismail2005}, employing fractional calculus, to construct the operator analytically. It is shown that this operator can be expressed in terms of the Weyl fractional integral, and allows one to overcome the difficulties associated with a non-local raising operator and efficiently compute the polynomial coefficients for all eigenstates. This provides an unexpected application of fractional calculus to analytically solvable potentials in Quantum Mechanics.

The rest of the paper is organized as follows. 
In Section \ref{sec:calc} the bound states for the Rosen-Morse potential are obtained in an analytical form, including their normalization constants. In Section \ref{sec:poschl} the easier case, when $\beta=0$, is treated, and it is shown that the ladder operators are local, are explicitly constructed in terms of derivatives, and allow the derivation of an efficient recurrence relation for Gegenbauer polynomials. 
In Section \ref{sec:frac} the general case is considered, and the construction of the raising operator is performed in two steps, by deriving a local recurrence relation between Jacobi polynomials applying the generating function formalism, and applying the Weyl fractional integral to perform an additional change in the polynomial parameters. In Section \ref{sec:concl}, the conclusions are summarized.

\section{Calculation of Rosen-Morse bound states}
\label{sec:calc}

In this section I present the solution for the Schr\"odinger equation for the Rosen-Morse potential \eqref{eq:rosen-morse}

\begin{equation}
-\frac{\hbar^2}{2m}\frac{d^2\psi}{dx^2} + V(x)\psi = E\psi,
\end{equation}
I shall follow closely the formalism of Refs. \onlinecite{Rosen1932} and \onlinecite{Nieto1978}. First of all, upon reparametrizing the equation with $x/\Delta \to x$ and $\Delta^2 E \to E$, we obtain the dimensionless form

\begin{equation}
-\psi'' -\alpha(\alpha+1)\sech^2 x\ \psi + 2\beta\tanh x\ \psi = E\psi.
\label{eq:schr}
\end{equation}

We are interested in bound state solutions ($E<-2\beta$). For this, it is convenient to write the wave function as

\begin{equation}
\psi(x) = e^{-ax}\cosh^{-b}x\ F(x).
\label{eq:subs}
\end{equation}

The motivation behind this expression is straight-forward. Since the $\sech^2$ term in \eqref{eq:schr} tends to 0 at $x\to\pm\infty$, the asymptotic behavior of the wavefunctions is completely determined by the $\tanh$ barrier. More specifically, for a bound state, we expect the wave function to decay exponentially. Since the energy barrier has different values at $x\to\pm\infty$, it is expected that the decay rate should be different at these two limits. This is precisely the behavior introduced by $e^{-ax}\cosh^{-b}x$, since at $x\to-\infty$ it falls like $ e^{(b-a)x}$ and at $x\to+\infty$ like $ e^{-(b+a)x}$. By carefully choosing $a$ and $b$, it should be possible to simplify \eqref{eq:schr}.

To perform the transformation, we write the second derivative of $\cosh^{-b} x$ as

\begin{equation}
\left(\cosh^{-b}x\right)'' = -b\left(\tanh x\cosh^{-b}x\right)' = [b^2-b(b+1)\sech^2 x]\cosh^{-b} x.
\end{equation}

From this expression, the second derivative of $e^{-ax}\cosh^{-b}x$ follows naturally:

\begin{equation}
\left(e^{-ax}\cosh^{-b}x\right)'' = [a^2+2ab\tanh x + b^2 - b(b+1)\sech^2 x]e^{-ax}\cosh^{-b}x.
\end{equation}

Rewriting \eqref{eq:schr} in the convenient form

\begin{equation}
\psi'' + [\alpha(\alpha+1)\sech^2 x -2\beta\tanh x + E]\psi = 0,
\end{equation}

and substituting \eqref{eq:subs}, we get

\begin{equation}
F'' - 2(a+b\tanh x)F' + \left\{[\alpha(\alpha+1)-b(b+1)]\sech^2 x + (2ab-2\beta)\tanh x + a^2+b^2 + E\right\}F = 0.
\label{eq:f}
\end{equation}

In order to find the most adequate values for $a$ and $b$, we must analyse the asymptotic behavior at $x\to\pm\infty$. For $x\to+\infty$, we have

\begin{equation}
F'' - 2(a+b)F' + [(a+b)^2 - 2\beta +E]F = 0,
\label{eq:pinf}
\end{equation}
and for $x\to-\infty$:
\begin{equation}
F'' - 2(a-b)F' + [(a-b)^2+2\beta+E]F = 0.
\label{eq:ninf}
\end{equation}

To simplify the equations, we choose $a$ and $b$ so that the last terms in \eqref{eq:pinf} and \eqref{eq:ninf} vanish. We choose

\begin{equation}
b + a = \sqrt{-E+2\beta}, \qquad b - a = \sqrt{-E-2\beta}.
\end{equation}

With this choice, we have two solutions at $x\to\infty$, the constant solution $F_1 \to C$ and $F_2 \propto e^{2(a+b)x}$, which propagate to $\psi$ as $\psi_1 \propto e^{-(a+b)x}$ and $\psi_2 \propto e^{(a+b)x}$. Since $b+a>0$, we are interested in solutions of \eqref{eq:f} which tend to a constant at $+\infty$.

In a completely analogous manner, the two solutions at $x\to-\infty$ are $F_1\to C$ and $F_2\propto e^{2(a-b)x}$, which propagate to $\psi_1 \propto e^{(b-a)x}$ and $\psi_2 \propto e^{-(b-a)x}$. Since $b-a>0$, the normalizable solution is, again, the one where $F$ tends to a constant at $-\infty$.

Therefore, we have

\begin{equation}
a = \frac{1}{2}\left(\sqrt{-E+2\beta}-\sqrt{-E-2\beta}\right), \qquad b = \frac{1}{2}\left(\sqrt{-E+2\beta}+\sqrt{-E-2\beta}\right).
\end{equation}

By substituting these into \eqref{eq:f}, many terms cancel and we can reduce it to the hypergeometric differential equation by performing the transformation

\begin{equation}
u = \frac{1}{2}\left(1+\tanh x\right), \qquad x = \frac{1}{2}\ln\left( \frac{u}{1-u}\right)
\end{equation}
and using the chain rule

\begin{equation}
\frac{dF}{dx} = 2u(1-u) \frac{dF}{du},
\end{equation}

\begin{equation}
\frac{d^2F}{dx^2} = \frac{d}{dx} \left(\frac{dF}{dx}\right) = 4u(1-u) \frac{d}{du}\left(u(1-u)\frac{dF}{du}\right) = 4u(1-u)(1-2u)\frac{dF}{du} + 4u^2(1-u)^2\frac{d^2F}{du^2}.
\end{equation}

The reduction to the hypergeometric differential equation is completed by rewriting \eqref{eq:f} in terms of the variable $u$, which implies $\tanh x = 2u-1$ and $\sech^2 x = 4u(1-u)$. After the dust settles down, we obtain

\begin{equation}
u(1-u)F'' + \left[b-a+1-2(b+1)u\right]F' + [\alpha(\alpha+1)-b(b+1)]F = 0,
\end{equation}
which is just the hypergeometric differential equation, conventionally written in the form

\begin{equation}
u(1-u)F'' + [t -(r+s+1)u]F' - rsF = 0,
\label{eq:hyper}
\end{equation}
where we have

\begin{equation}
r = b-\alpha, \qquad s = b+\alpha+1, \qquad t = b-a+1.
\end{equation}

We are interested in solutions at the interval $0<u<1$, which are not singular at the extremes $u=0$ and $u=1$. The solution of \eqref{eq:hyper} which is finite at $u=0$ is simply the hypergeometric function

\begin{equation}
F(r,s;t;u) = \frac{\Gamma(t)}{\Gamma(r)\Gamma(s)}\sum_{n=0}^\infty \frac{\Gamma(r+n)\Gamma(s+n)}{\Gamma(t+n)} \frac{u^n}{n!}.
\end{equation}

To understand the behavior at $u\to 1$, we use the linear transformation formula\cite{Abramowitz1964}:

\begin{align}
F(r,s;t;u) &= \frac{\Gamma(t)\Gamma(t-r-s)}{\Gamma(t-r)\Gamma(t-s)}F(r,s;r+s-t+1;1-u) \nonumber  \\
 &+ (1-u)^{t-r-s}\frac{\Gamma(t)\Gamma(r+s-t)}{\Gamma(r)\Gamma(s)}F(t-r,t-s;t-r-s+1;1-u).
\label{eq:trans}
\end{align}

From the definition of the hypergeometric function, we have $F(r,s;t;0)=1$. Since we have $t-r-s=-(b+a)<0$, the second term in \eqref{eq:trans} diverges as $u\to1$. Therefore, it needs to vanish so that we obtain a normalizable state. That happens when the Gamma functions at the denominators are at the poles, which occur at nonpositive integral values of the arguments. Thus, we have the condition

\begin{equation}
r = b - \alpha = -n, \qquad n=0,1,\ldots,
\end{equation}
from which the energy of each bound state follows:

\begin{equation}
E_n =  - (\alpha-n)^2 -\frac{\beta^2}{(\alpha-n)^2}.
\end{equation}

The highest possible value of $n$ is given by the condition $b-a>0$, which implies

\begin{equation}
\alpha - n > \frac{\beta}{\alpha-n} \Rightarrow n < \alpha - \sqrt{\beta}.
\end{equation}

The wave functions can be written in an elegant manner in terms of Jacobi polynomials. Using the definition\cite{Abramowitz1964}

\begin{equation}
P^{\alpha,\beta}_n(x) = \frac{\Gamma(n+\alpha+1)}{\Gamma(n+1)\Gamma(\alpha+1)} F(-n,n+\alpha+\beta+1;\alpha+1;(1-x)/2),
\end{equation}
and also the relation

\begin{equation}
P^{\alpha,\beta}_n(-x) = (-1)^nP^{\beta,\alpha}_n(x),
\end{equation}
we derive

\begin{equation}
F(-n,n+2b+1,b-a+1;u) = (-1)^n \frac{\Gamma(n+1)\Gamma(b-a+1)}{\Gamma(n+b-a+1)} P^{b+a,b-a}_n(2u-1).
\end{equation}

Therefore, it is possible to express the bound state solutions as

\begin{equation}
\psi_n(x) = A_n e^{-ax}\cosh^{-b}(x) P^{b+a,b-a}_n(\tanh x).
\end{equation}

The normalization constant $A_n$ is obtained by performing the change of variable $v=\tanh x$ and writing the integral

\begin{equation}
|A_n|^{-2} = \int_{-1}^1 (1-v)^{b+a-1}(1+v)^{b-a-1}P^{b+a,b-a}_n(v)^2dv,
\label{eq:norm}
\end{equation}
which can be evaluated using the identity\cite{Bateman1954}

\begin{equation}
\int_{-1}^1 (1-v)^{\alpha-1}(1+v)^\beta P^{\alpha,\beta}_n(v)^2 dv = \frac{2^{\alpha+\beta}\Gamma(\alpha+n+1)\Gamma(\beta+n+1)\Gamma(\alpha)}{n!\Gamma(\alpha+1)\Gamma(\alpha+\beta+n+1)}.
\label{eq:it1}
\end{equation}

Changing variables from $v\to-v$ and using $P_n^{\alpha,\beta}(v) = (-1)^nP_n^{\beta,\alpha}$, the following identity follows:

\begin{equation}
\int_{-1}^1 (1-v)^{\alpha}(1+v)^{\beta-1} P^{\alpha,\beta}_n(v)^2 dv = \frac{2^{\alpha+\beta}\Gamma(\alpha+n+1)\Gamma(\beta+n+1)\Gamma(\beta)}{n!\Gamma(\beta+1)\Gamma(\alpha+\beta+n+1)}.
\label{eq:it2}
\end{equation}

The integral at \eqref{eq:norm} can be reduced to a sum of \eqref{eq:it1} and \eqref{eq:it2} by inserting into the integrand

Our integral can be reduced to these two by inserting into the integrand

\begin{equation}
1 = \frac{1}{2}\left[(1+v)+(1-v)\right],
\end{equation}
and we have

\begin{equation}
|A_n|^{-2} = \frac{2^{2b-1}\Gamma(b+a+n+1)\Gamma(b-a+n+1)}{n!\Gamma(2b+n+1)}\left(\frac{1}{b+a}+\frac{1}{b-a}\right).
\end{equation}

Now, we can write $b$ and $a$ in terms of $n$:

\begin{equation}
b = \alpha - n, \qquad a = \frac{\beta}{(\alpha-n)}
\end{equation}

Thus:

\begin{equation}
|A_n|^{-2} = \frac{2^{2\alpha-2n}\Gamma\left(\alpha+\beta/(\alpha-n)+1\right)\Gamma\left(\alpha-\beta/(\alpha-n)+1\right)(\alpha-n)^3}{n!\Gamma(2\alpha-n+1)[(\alpha-n)^4-\beta^2]}
\end{equation}

Many conventions for the phase of $A_n$ are possible. In this work, I choose $A_n$ as a positive real number for all bound states. After all calculations, we can finally express the wave functions for the bound states as

\begin{equation}
\psi_n(x) = A_ne^{-\beta x/(\alpha-n)}\sech^{(\alpha-n)}x P_n^{\alpha-n+\beta/(\alpha-n),\alpha-n-\beta/(\alpha-n)}(\tanh x).
\label{eq:wave}
\end{equation}

\section{Ladder operators and the efficient computation of wave functions}
\label{sec:poschl}

Even with the wave functions written in closed form as in \eqref{eq:wave}, for practical applications, it is often necessary to evaluate $\psi_n$ at a given value $x$. In order to do that, a procedure to determine the Jacobi polynomials in \eqref{eq:wave} is necessary. The standard way to do this is to apply the two-term recurrence relation\cite{Gradshteyn2014}

\begin{align}
2(n+1) & (n+\alpha+\beta+1)(2n+\alpha+\beta)P_{n+1}^{\alpha,\beta}(x) = \nonumber \\
& (2n+\alpha+\beta+1)[(2n+\alpha+\beta)(2n+\alpha+\beta+2)x+\alpha^2-\beta^2]P_n^{\alpha,\beta}(x)
 -\\  & 2(n+\alpha)(n+\beta)(2n+\alpha+\beta+2)P_{n-1}^{\alpha,\beta}(x), \nonumber
\end{align}
and evaluate the coefficients. This solution is not ideal, however, especially when the number of bound states is very large. It is readily noticed that the parameters for the Jacobi polynomials in \eqref{eq:wave} change depending on the number $n$ of the bound state, while they remain fixed in the previous recurrence relation. That means that, for states with very large $n$, one must calculate the coefficients of Jacobi polynomials all the way down, which are unrelated to the previous bound states.

In this section, it will be shown how the computation of the polynomial coefficients becomes much more efficient when raising and lowering operators are written explicitly. In this manner, it is possible to find a recurrence relation between the coefficients of the $(n+1)$-th bound state in terms of the $n$-th one, and thus only the coefficients relevant for the evaluation of the wave functions are computed.

We shall work first with the symmetric ($\beta=0$) case to illustrate the method. In this case, the Rosen-Morse potential reduces to the modified P\"oschl-Teller one. In the notation from the previous section, we find $a=0$ and the bound state energies reduce to the well known formula

\begin{equation}
E_n = -(\alpha-n)^2.
\end{equation}

The two parameters for the Jacobi polynomials in \eqref{eq:wave} become identical, and the solution can be expressed in terms of Gegenbauer polynomials\cite{Abramowitz1964}

\begin{equation}
C_n^{\lambda}(v) = \frac{\Gamma(\lambda+\frac{1}{2})\Gamma(2\lambda+n)}{\Gamma(2\lambda)\Gamma(\lambda+n+\frac{1}{2})}P_n^{\lambda-\frac{1}{2},\lambda-\frac{1}{2}}(v),
\end{equation}
this identity is equivalent to

\begin{equation}
P_n^{\alpha-n,\alpha-n}(v) =  \frac{\Gamma(2\alpha-2n+1)\Gamma(\alpha+1)}{\Gamma(\alpha-n+1)\Gamma(2\alpha-n+1)} C_n^{\alpha-n+\frac{1}{2}}(v).
\end{equation}

Applying this, the wave functions for the modified P\"oschl-Teller potential are simply

\begin{equation}
\psi_n(x) = \frac{2^{n-\alpha}\Gamma(2\alpha-2n+1)}{\Gamma(\alpha-n+1)}\sqrt{\frac{n!(\alpha-n)}{\Gamma(2\alpha-n+1)}}\sech^{(\alpha-n)}x C_n^{\alpha-n+1/2}(\tanh x).
\end{equation}

The problem of the efficient evaluation of the polynomial coefficients continues. Since the parameter for the Gegenbauer polynomials also depends on $n$, the application of the two-term recurrence

\begin{equation}
(n+2)C_{n+2}^\lambda(v) = 2(\lambda+n+1)vC_{n+1}^\lambda(v) - (2\lambda+n)C_n^\lambda(v),
\end{equation}
also requires the computation of coefficients unrelated to any bound state.

Fortunately, the problem disappears when we write the solution instead as a Legendre function. If we consider the so-called associated Legendre functions on the cut (those defined for $-1<x<1$), defined as\cite{Gradshteyn2014}

\begin{equation}
P_\nu^\mu (x) = 	\frac{1}{\Gamma(1-\mu)}\left(\frac{1+x}{1-x}\right)^{\frac{\mu}{2}} F(-\nu,\nu+1;1-\mu;(1-x)/2),
\end{equation}
we can relate them to Gegenbauer polynomials by means of the identity\cite{Abramowitz1964}

\begin{equation}
C_n^\lambda(x) = \frac{\Gamma(2\lambda+n)}{n!\Gamma(2\lambda)}F(-n,n+2\lambda;\lambda+1/2;(1-x)/2).
\end{equation}

With the parametrization used in this work, that becomes

\begin{equation}
C_n^{\alpha-n+\frac{1}{2}}(v) = \frac{\Gamma(2\alpha-n+1)}{n!\Gamma(2\alpha-2n+1)} F(-n,-n+2\alpha+1;\alpha-n+1;(1-v)/2).
\end{equation}

Furthermore, if we apply the transformation identity\citep{Abramowitz1964}

\begin{equation}
F(r,s;t;(1-x)/2) = \left(\frac{1+x}{2}\right)^{t-r-s}F(t-r,t-s;t;(1-x)/2),
\end{equation}
it is possible to write

\begin{equation}
C_n^{\alpha-n+\frac{1}{2}}(v) = \frac{\Gamma(2\alpha-n+1)}{n!\Gamma(2\alpha-2n+1)}\left(\frac{1+v}{2}\right)^{n-\alpha}F(-\alpha,\alpha+1;\alpha-n+1;(1-v)/2),
\end{equation}
and we can represent the wavefunctions in terms of the associated Legendre functions by recognizing $\nu=\alpha$ and $\mu=n-\alpha$:

\begin{equation}
F(-\alpha,\alpha+1;\alpha-n+1;(1-v)/2) = \Gamma(\alpha-n+1) \left(\frac{1-v}{1+v}\right)^\frac{n-\alpha}{2}P^{n-\alpha}_\alpha(v).
\end{equation}

The final answer is

\begin{equation}
\psi_n(x) = \sqrt{\frac{(\alpha-n)\Gamma(2\alpha-n+1)}{n!}}
P_\alpha^{n-\alpha}(\tanh x).
\end{equation}

The ladder operators for the modified P\"oschl-Teller potential arise from the identities\cite{Gradshteyn2014}

\begin{equation}
(1-x^2)\frac{dP_\nu^\mu(x)}{dx} = -\sqrt{1-x^2}P_\nu^{\mu+1}(x) -\mu xP_\nu^\mu(x) = (\nu-\mu+1)(\nu+\mu)\sqrt{1-x^2}P_\nu^{\mu-1}(x)+\mu xP_\nu^\mu (x).
\end{equation}

Rearranging, we find

\begin{equation}
P^{\mu+1}_\nu(x) = \left[-\sqrt{1-x^2}\frac{d}{dx} - \frac{\mu x}{\sqrt{1-x^2}}\right]P^\mu_\nu(x), \qquad P^{\mu-1}_\nu(x) = \frac{\left[\sqrt{1-x^2}\frac{d}{dx}-\frac{\mu x}{\sqrt{1-x^2}}\right]}{(\nu-\mu+1)(\nu+\mu)}P^\mu_\nu(x).
\end{equation}

Applying the first operator in brackets to the wave functions $\psi_n$ and differentiating with respect to $v=\tanh x$, we obtain

\begin{equation}
\left[-\sqrt{1-v^2}\frac{d}{dv} + \frac{(\alpha-n)v}{\sqrt{1-v^2}}\right]\psi_n(v) =  \sqrt{\frac{(\alpha-n)\Gamma(2\alpha-n+1)}{n!}}
P_\alpha^{n+1-\alpha}(v).
\end{equation}

It is clear that this operator can be identified as a creation operator $a^\dagger$. After some straightforward algebraic manipulations, we find

\begin{equation}
a^\dagger|n\rangle = \sqrt{\frac{(n+1)(2\alpha-n)(\alpha-n)}{\alpha-n-1}}|n+1\rangle.
\end{equation}

Its hermitian conjugate is obtained by flipping the sign of the derivative:

\begin{equation}
\left[\sqrt{1-v^2}\frac{d}{dv} + \frac{(\alpha-n)v}{\sqrt{1-v^2}}\right]\psi_n(v) =  n(2\alpha-n+1)\sqrt{\frac{(\alpha-n)\Gamma(2\alpha-n+1)}{n!}}
P_\alpha^{n-1-\alpha}(v).
\end{equation}

We find

\begin{equation}
a|n\rangle = \sqrt{\frac{n(2\alpha-n+1)(\alpha-n)}{\alpha-n+1}}|n-1\rangle,
\end{equation}
which implies

\begin{equation}
a^\dagger a|n\rangle = n(2\alpha-n+1)|n\rangle, \qquad aa^\dagger|n\rangle = (n+1)(2\alpha-n)|n\rangle, \qquad [a,a^\dagger]|n\rangle = 2(\alpha-n)|n\rangle.
\end{equation}

It has been noted elsewhere\cite{Dong2002} that, by defining the operator $a_0|n\rangle = (\alpha-n)|n\rangle$, one arrives at an SU(2) algebra. Here we are concerned with the efficient evaluation of the wave functions. In any case, we write the operators in terms of the $x$ variable as

\begin{equation}
a^\pm|n\rangle = \left[\mp \cosh x \frac{d}{dx} + (\alpha-n)\sinh x \right]\psi_n(x).
\end{equation}

The raising operator $a^\dagger$ allows us to obtain an efficient recurrence relation for the polynomial coefficients of the wave functions. To do this, we write the $n-th$ state wave function as

\begin{equation}
\psi_n(x) = \sum_{m=0}^n a_{mn}\sech^{(\alpha-n)}x\tanh^m x,
\end{equation}
where it should be clear that

\begin{equation}
a_{00} = \frac{2^{-\alpha}}{\Gamma(\alpha+1)}\sqrt{\alpha\Gamma(2\alpha+1)}.
\end{equation}

Applying the raising operator

\begin{equation}
\sum_{m=0}^{n+1}a_{m,n+1}\tanh^m x = \sqrt{\frac{\alpha-n-1}{(n+1)(2\alpha-n)(\alpha-n)}}\sum_{m=0}^n a_{mn}\left[(2\alpha-2n+m)\tanh^{m+1}x - m \tanh^{m-1}x\right],
\end{equation}
and changing the summation variable we obtain

\begin{equation}
a_{m,n+1} = \sqrt{\frac{\alpha-n-1}{(n+1)(2\alpha-n)(\alpha-n)}}\left[(2\alpha-2n+m-1)a_{m-1,n} - (m+1)a_{m+1,n}\right].
\end{equation}

From this relation, only the polynomial coefficients relevant to the computation of the bound state wave functions are evaluated.

\section{Fractional calculus and the Rosen-Morse raising operator}
\label{sec:frac}

We now proceed to determine a similar recurrence relation for eigenstates of the non-symmetric case ($\beta\neq0$). It is immediately noticed that we must find a relation between the Jacobi polynomials whose parameters vary by non-integer values. This presents a serious problems, since identities involving derivatives of the hypergeometric function only change its arguments in integral steps. Therefore, it must be concluded that the raising operator for general Rosen-Morse eigenstates should be non-local. Indeed, it has been noted by Ismail\cite{Ismail2005} that it is possible to change the parameters of Jacobi polynomials with the relation

\begin{equation}
x^{-\lambda-\alpha}I_0^\lambda\left[x^\alpha P_n^{\alpha,\beta}(1-2x)\right] = \frac{\Gamma(\alpha+n+1)}{\Gamma(\alpha+\lambda+n+1)}P_n^{\alpha+\lambda,\beta-\lambda}(1-2x),
\end{equation}
where the operator $I^\lambda_a$ is the Riemann-Liouville fractional integral, defined by

\begin{equation}
I_a^\lambda[f(x)] = \int_a^x \frac{(x-t)^{\lambda-1}}{\Gamma(\lambda)}f(t)dt.
\end{equation}

To construct the raising operator, we shall need the adjoint of $I_0^\lambda$. Since the hypergemetric function is defined on $0<x<1$, it is given by the Weyl fractional integral, defined as

\begin{equation}
W^\nu[f(x)] = \int_x^1 \frac{(t-x)^{\nu-1}}{\Gamma(\nu)}f(t)dt.
\label{eq:weyl}
\end{equation}

But first, it shall be necessary to find a relation between $P_n^{\alpha,\beta}$ and $P_{n+1}^{\alpha-1,\beta-1}$. This can be done in a concise manner applying the formalism of generating functions\cite{Wilf1990}. First, we define the generating function

\begin{equation}
F(x,s) = \sum_{n=0}^\infty P_n^{\alpha-n,\beta-n}(x)s^n,
\end{equation}
and write the Jacobi polynomials using the expansion\cite{Gradshteyn2014}

\begin{equation}
P_n^{\alpha-n,\beta-n}(x) = \frac{1}{2^n}\sum_{m=0}^n \binom{\alpha}{m}\binom{\beta}{n-m}(x-1)^{n-m}(x+1)^m.
\end{equation}

After reindexing the sum, we get

\begin{equation}
F(x,s) = \sum_{n,m=0}^\infty \binom{\alpha}{m}\binom{\beta}{n}\left(\frac{x-1}{2}s\right)^{n}\left(\frac{x+1}{2}s\right)^m,
\end{equation}
and recognize the generalized binomial expansion. Therefore, it is possible to write the generating function in closed form:

\begin{equation}
F(x,s) = \left[1+\frac{(x+1)s}{2}\right]^\alpha \left[1+\frac{(x-1)s}{2}\right]^\beta.
\end{equation}

Now we expect to obtain a recurrence relation for the polynomials if we find a differential equation for $F(x,s)$. To do this, write the derivative as

\begin{equation}
\frac{\partial F}{\partial s} = \frac{\alpha(x+1)}{2}\left[1+\frac{(x+1)s}{2}\right]^{\alpha-1}\left[1+\frac{(x-1)s}{2}\right]^\beta + \frac{\beta(x-1)}{2} \left[1+\frac{(x+1)s}{2}\right]^\alpha \left[1+\frac{(x-1)s}{2}\right]^{\beta-1}.
\end{equation}

After some algebraic manipulations, we obtain the differential equation

\begin{align}
\left[\left(1+\frac{xs}{2}\right)^2-\frac{s^2}{4}\right]\frac{\partial F}{\partial s} &= \frac{F}{2}\left\{\alpha(x+1)\left[1+\frac{(x-1)s}{2}\right]+\beta(x-1)\left[1+\frac{(x+1)s}{2}\right]\right\}, \\
\left[1 + sx + \frac{s^2}{4}(x^2-1)\right] \frac{\partial F}{\partial s}&= \frac{F}{2}\left[\alpha(x+1)+\beta(x-1) + s(x^2-1)\frac{\alpha+\beta}{2}\right].
\end{align}

Substituting the Taylor expansion in $s$ (we denote the polynomials by $P_n$ temporarily):

\begin{equation}
\sum_{n=0}^\infty nP_n(x)s^{n-1} +nxP_n(x)s^n +\frac{n}{4}(x^2-1)P_n(x)s^{n+1} = \sum_{n=0}^\infty \left[\alpha(x+1)+\beta(x-1)\right]\frac{P_n(x)}{2}s^n + (\alpha+\beta)(x^2-1) \frac{P_n(x)}{4}s^{n+1}
\end{equation}

From this expression, the recurrence relation finally appears after reindexing the summation:

\begin{equation}
(n+1)P_{n+1}(x) = \left[\frac{\alpha(x+1)+\beta(x-1)-2nx}{2}\right]P_n(x) + \frac{\alpha+\beta-n+1}{4}(x^2-1)P_{n-1}(x).
\end{equation}

It is possible to simplify further using\cite{Gradshteyn2014}

\begin{equation}
\frac{dP_n^{\alpha-n,\beta-n}(x)}{dx} = \frac{\alpha+\beta-n+1}{2} P_{n-1}^{\alpha-n+1,\beta-n+1}(x),
\end{equation}
and we finally obtain

\begin{equation}
2(n+1)P_{n+1}^{\alpha-n-1,\beta-n-1}(x) = \left[-(1-x^2)\frac{d}{dx} + (\alpha+\beta-2n)x + (\alpha - \beta) \right]P^{\alpha-n,\beta-n}_n(x),
\end{equation}
or the more useful form

\begin{equation}
2(n+1)P_{n+1}^{\alpha-1,\beta-1}(x) = \left[-(1-x^2)\frac{d}{dx} + (\alpha+\beta)x + (\alpha - \beta) \right]P^{\alpha,\beta}_n(x).
\label{eq:recjac}
\end{equation}

To make further progress, it will be necessary to understand the effect of the convolution integral \eqref{eq:weyl} on polynomials. It will be convenient to expand the polynomials in coefficients $a_m$ as

\begin{equation}
f(x) = \sum_{m=0}^n a_m(1-x)^m.
\end{equation}

Therefore, since the operator \eqref{eq:weyl} is linear, it suffices to understand its effect on functions of the form $(1-x)^\rho$. We have

\begin{equation}
W^\nu[(1-x)^\rho] = \frac{1}{\Gamma(\nu)}\int_x^1 (t-x)^{\nu-1}(1-t)^\rho dt,
\end{equation}
and the integral reduces to the beta function when performing the change $t=x + u(1-x)$:

\begin{equation}
W^\nu[(1-x)^\rho] = \frac{(1-x)^{\nu+\rho}}{\Gamma(\nu)} \int_0^1 u^{\nu-1}(1-u)^\rho du = \frac{\Gamma(\rho+1)}{\Gamma(\rho+\nu+1)} (1-x)^{\nu+\rho}.
\end{equation}

To see how $W^\nu$ changes the parameters of Jacobi polynomials, we should evaluate

\begin{equation}
G(x) = W^\nu[(1-x)^\alpha P_n^{\alpha,\beta}(x)].
\end{equation}

To do this, we first write the Jacobi polynomials in terms of hypergeometric functions:

\begin{equation}
P_n^{\alpha,\beta}(x) = \frac{\Gamma(\alpha+n+1)}{n!\Gamma(\alpha+1)}F(-n,n+\alpha+\beta+1;1+\alpha;(1-x)/2).
\end{equation}

The calculation is straightforward:

\begin{align}
G(x) &= \frac{\Gamma(\alpha+n+1)}{n!\Gamma(\alpha+1)}I^\nu\left[\sum_{k=0}^n \frac{(-n)_k(n+\alpha+\beta+1)_k}{2^kk!(\alpha+1)_k}(1-x)^{k+\alpha}\right] \\
&= \frac{\Gamma(\alpha+n+1)}{n!\Gamma(\alpha+1)} \sum_{k=0}^n \frac{(-n)_k(n+\alpha+\beta+1)_k}{2^kk!(\alpha+1)_k} \frac{\Gamma(\alpha+k+1)}{\Gamma(\alpha+k+\nu+1)} (1-x)^{k+\alpha+\nu} \\
&= \frac{(1-x)^{\alpha+\nu}\Gamma(\alpha+n+1)}{n!\Gamma(\alpha+\nu+1)} \sum_{k=0}^n \frac{(-n)_k(n+\alpha+\beta+1)_k}{2^kk!(1+\alpha+\nu)_k}(1-x)^k \\
&= \frac{(1-x)^{\alpha+\nu}\Gamma(\alpha+n+1)}{n!\Gamma(\alpha+\nu+1)} F(-n,n+\alpha+\beta+1;1+\alpha+\nu;(1-x)/2).
\end{align}

Using the fact that

\begin{equation}
P_n^{\alpha+\nu,\beta-\nu}(x) = \frac{\Gamma(\alpha+\nu+n+1)}{n!\Gamma(\alpha+\nu+1)}F(-n,n+\alpha+\beta+1;1+\alpha+\nu;(1-x)/2),
\end{equation}
we obtain

\begin{equation}
G(x) = \frac{(1-x)^{\alpha+\nu}\Gamma(\alpha+n+1)}{\Gamma(\alpha+\nu+n+1)}P_n^{\alpha+\nu,\beta-\nu}(x).
\label{eq:conv}
\end{equation}

From this, we have the relation

\begin{equation}
P_{n+1}^{\alpha-1+\nu,\beta-1-\nu}(x) = \frac{\Gamma(\alpha+\nu+n+1)}{2(n+1)\Gamma(\alpha+n+1)}(1-x)^{-\alpha-\nu}I^\nu\left\{(1-x)^\alpha\left[-(1-x^2)\frac{d}{dx} + (\alpha+\beta-2n)x + (\alpha - \beta)\right]P_n^{\alpha,\beta}(x)\right\},
\end{equation}
which can be used to construct a raising operator for Rosen-Morse bound states. Express the Jacobi polynomials as

\begin{equation}
P^{\alpha,\beta}_n(x) = \sum_{m=0}^n a_m(1-x)^m, \qquad P^{\alpha-1,\beta-1}_{n+1}(x) = \sum_{m=0}^{n+1} b_m(1-x)^m, \qquad P^{\alpha-1+\nu,\beta-1-\nu}_{n+1}(x) = \sum_{m=0}^{n+1} c_m(1-x)^m.
\end{equation}

By applying this expansion on \eqref{eq:recjac}, we obtain the following relation between the polynomial coefficients:

\begin{equation}
2(n+1)b_m = -(\alpha+\beta+m-1)a_{m-1} + 2(\alpha+m)a_m.
\label{eq:ba}
\end{equation}

Finally, if we apply the convolution identity \eqref{eq:conv} with $\nu=\frac{\beta}{(\alpha-n-1)(\alpha-n)}$, we will obtain the Jacobi polynomial for the $(n+1)$-th state in terms of the $n$-th one:

\begin{equation}
c_m = \frac{\Gamma(\alpha+\nu+n+1)\Gamma(\alpha+m)}{\Gamma(\alpha+n+1)\Gamma(\alpha+m+\nu)}b_m.
\label{eq:cb}
\end{equation}

\begin{figure}[h!]
\includegraphics{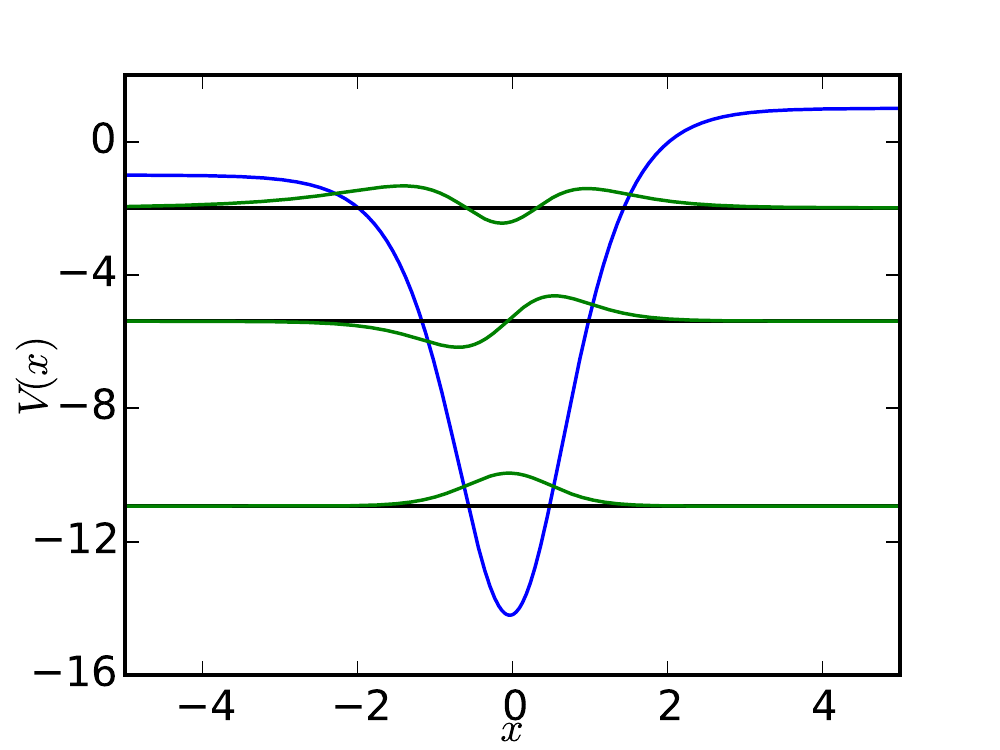}
\caption{\label{fig:states}(Color online) Rosen-Morse potential (blue line) with parameters $\alpha=3.3$ and $\beta=0.5$. For this choice, the potential admits three bound states with energies denoted by the black lines. The wave functions (green lines) are expressed in terms of Jacobi polynomials, computed according to the procedure outlined in the text. They have the expected form for bound states confined in a well, with the number of nodes increasing with state number, and part of the wave function leaking into the classically forbidden region.}
\end{figure}

The procedure completes the construction of the raising operator applying the Weyl fractional integral and also provides an efficient algorithm for evaluating the Rosen-Morse wave functions. The Jacobi polynomial corresponding to the ground state is simply $P^{\alpha+\beta/\alpha,\alpha-\beta/\alpha}_0(x)=1$. In order to raise the $n$-th bound state to the $(n+1)$-th, we apply formulas \eqref{eq:ba} and \eqref{eq:cb} with the substitutions

\begin{equation}
\alpha \to \alpha - n + \beta/(\alpha-n), \qquad \beta \to \alpha - n - \beta/(\alpha-n), \qquad \nu=\frac{\beta}{(\alpha-n-1)(\alpha-n)},
\end{equation}
and with the polynomial coefficients computed, it is trivial to obtain the value of the wave function at any $x$. An example of such calculation is provided in Figure~\ref{fig:states}.

\section{Conclusions}
\label{sec:concl}

In this paper, a new recurrence relation between Jacobi polynomials arising from the solution of the Schr\"odinger equation for the Rosen-Morse potential was introduced. The formulae provide an efficient way to numerically evaluate the bound state wave functions, and the techniques used to derive them may be adapted to study other related potentials, such as the trigonometric Rosen-Morse and Eckhart potentials.

It was shown that the construction of ladder operators for the general potential is fundamentally different from the one used in the symmetric case, which is generally known as the modified P\"oschl-Teller potential. The key difference relies on the fact that the ladder operators should be non-local and need to be defined in terms of convolution integrals typical of fractional calculus, namely the Weyl fractional integral. This result provides a novel application of fractional calculus to analytically solvable quantum systems, invinting more research into the applicability of similar techniques to more general exactly solvable potentials.

\subsection*{Acknowledgements} 
The author acknowledges fruitful discussions with Sergue\"i Tchoumakov and thanks CAPES (PVE grant no. 88887.116797/2016-00) for financial support.



%

\end{document}